\documentclass[twocolumn]{aastex63}

\received{}
\revised{\today}
\accepted{}

\submitjournal{ApJL}

\shorttitle{Is 2I/Borisov a Stardust Comet?}
\shortauthors{Eubanks}
\usepackage{gensymb} 
\usepackage{multirow}
\usepackage{amsmath}

\begin{document}

\title{Is Interstellar Object 2I/Borisov a Stardust Comet? 
\\Predictions for the Post Perihelion Period}

\correspondingauthor{T. Marshall Eubanks}
\email{tme@space-initiatives.com}
\author[0000-0001-9543-0414]{T.M. Eubanks}
\affil{Space Initiatives Inc and I4IS (US),
\\Newport, Virginia 20124}



%% Mark off your abstract in the ``abstract'' environment.

\begin{abstract}
The detection of interstellar bodies passing near the Sun offers the opportunity to 
observe not just objects similar to those in the solar system, but also unfamiliar objects 
without solar system analogues. 
Here I show that Asymptotic Giant Branch (AGB) stellar evolution may lead to the creation, out of stardust, of substantial numbers of nomadic Post-Main-Sequence Objects (PMSOs). AGB nucleosynthesis 
will produce three broad classes of PMSO chemistry, oxygen, carbon and nitrogen rich (O-rich, C-rich, N-rich, respectively), depending largely on the original stellar mass.
I further show that the Interstellar Comet 2I/Borisov (2I)  belongs to a kinematic dynamical stream, the Wolf 630 stream, with an age and galactic orbit consistent with its origination as a stardust comet; the apparent lack of water
in the 2I coma is consistent with it  being a C-rich PMSO. 
I also provide predictions for distinguishing stardust comets from more conventional 
interstellar comets and asteroids ejected during planetary formation; these can be applied to 2I  
in its upcoming observational phase in early 2020 as it moves away from the Sun. In particular, isotope ratios 
of the CNO elements could be dispositive, 
IR detection of the 11.3 $\mu$m SiC line, the 30 $\mu$m line, or the IR PAH lines would provide strong evidence for a C-rich PMSO and detection of Na or Li enhancement would indicate an N-rich PMSO. 
\end{abstract}




\section{Introduction}

The chemical evolution of the galaxy is dominated by elements created by 
slow neutron capture (the s-process)  in  AGB stars, 
\citep{Busso-et-al-1999-a},
and by the rapid processes present
in explosive nucleosynthesis in Supernova (SN) \citep{Yoshida-2007-a}.
The isotopic and elemental composition of presolar grains in primitive meteorites 
shows that they are predominately from these two sources \citep{Davis-2011-a}, and half 
of all the elements heavier than iron are thought to result from the s-process \citep{Herwig-2005-a}.
AGB material is thus very
important in the chemical evolution of the galaxy, and its presence in stardust seems to 
permeate the galactic disk \citep{Gail-et-al-2009-a}. 
Although PMSOs, condensations of AGB gas and dust, could thus be very common in the galaxy, they will not be present in the solar system except 
as passing ISOs; every newly found ISO should thus be considered to be a possible PMSO.

The discovery, on August 30, 2019 by the amateur astronomer Gennadiy Borisov,  of the interstellar comet 2I (originally C/2019 Q4 (Borisov)) \citep{Guzik-et-al-2019-a} provides a second InterStellar Object (ISO) available for study in its passage through our solar system. 2I was rapidly recognized to be both on a hyperbolic orbit and an active object, superficially similar to a small solar system comet \citep{Jewitt-and-Luu-2019-a} with an estimated nucleus
diameter of order 1 km \citep{Jewitt-et-al-2019-a}.
Unlike the case of the first known ISO, 1I/'Oumuamua (or 1I), which never displayed any cometary 
activity or detectable spectral line emissions \citep{Trilling-et-al-2018-b}, 2I is 
an active comet well situated for a spectroscopic observing campaign 
in early 2020, after its perihelion and closest approach to the Earth in December, 2019.

\section{Formation of Small Objects After the Stellar Main Sequence}
\label{Sec:Post-Main-Sequence-Object}

Table \ref{table:Predictions} shows, in its top line, the dependence of AGB chemistry on mass (for 
an introduction to this complicated subject, see 
\citep{Herwig-2005-a,Busso-et-al-1999-a,Gail-et-al-2009-a,Ventura-et-al-2017-a,Ventura-et-al-2018-a}). 
Stars with masses, M, between $\sim$0.8 M$_{\bigodot}$ and 8 M$_{\bigodot}$ (where M$_{\bigodot}$  is the mass of the the Sun) will, after they leave the Main Sequence, 
pass through a Red Giant stage burning He in their core. Once core burning
is completed they pass to their AGB stage, 
where they are powered by thin shells burning H and He above a largely inert C and O core. 

In an AGB star H burning creates He over time. When a sufficient amount has accumulated, 
He burning ignites, causing a thermal pulse, which may last a few centuries. This causes
strong convection which carries processed shell material to the stellar surface, the so-called
 "third dredge up." At the end of the shell H burning
 the star will shed even more material, becoming very hot and possibly causing a Planetary Nebulae from ionized gas emissions \citep{Ventura-et-al-2018-a}. Once the outer envelope is fully expelled, the remaining hot core  becomes a white dwarf and the ejected material cools rapidly. 

Although most stars have O-rich surfaces during their period on the main sequence, 
in the AGB stage each thermal pulse brings more carbon to the surface.
For stars with M $\lesssim$  1.5 M$_{\bigodot}$, there are not enough of these events
to create a C-rich star. More massive stars, up to roughly 3 - 3.5 M$_{\bigodot}$, 
dredge up more and material from their interior and become C-rich stars with C/O~$>$~1, while 
AGB stars with M $\gtrsim$ 3 - 3.5 M$_{\bigodot}$ undergo Hot Bottom Burning (HBB) and exhibit the signatures of proton-capture (CNO-chain) nucleosynthesis in their ejected gas and dust \citep{Ventura-et-al-2018-a}. This drives the C-N-O abundances to a steady state where $^{14}$N dominates, leading to a N-rich stellar surface.
The consumption of $^{12}$C in this process will lead to  very low
$^{12}$C/$^{13}$C ratios and a modified O-rich chemistry in the ejecta. 
HBB burning of He will create Lithium through  the fusion chain
$^{3}$He +  $^{4}$He $\rightarrow$ $^{7}$Be $\rightarrow$ $^{7}$Li
\citep{Ventura-et-al-2018-a}, leading to the possibility of Lithium rich PMSO.

AGB and post-AGB stars eject a substantial amount of their total mass as dust and gas. These dust driven winds are
subject to instabilities
\citep{Hopkins-et-al-2018-a}, leading to the molecular condensations visible in PNe, where
they are known as cometary knots \citep{Odell-et-al-2003-a,Odell-et-al-2007-a,Matsuura-et-al-2009-a}. These condensations, which have an individual mass comparable to an Earth mass and a total mass of $\sim$0.2  M$_{\bigodot}$ for the relatively nearby Helix Nebula (NGC 7293) \citep{Huggins-et-al-2006-a}, appear to be suitable locations for the condensation of PMSO during and after the PN stage. As these condensations are not gravitationally bound,
with relative velocities of $\sim$10 km s$^{-1}$ to their host stars, they could release large numbers of nomadic PMSO into the galaxy, potentially including planetary mass bodies \citep{Odell-et-al-2003-a}. 

There appears to have been little previous discussion of the possibility of PMSO in the galaxy, although there has been consideration of both C-rich and N-rich planets forming from AGB and SN ejecta  \citep{Kuchnet-Seager-2005-a}. 
C-rich planets may exist \citep{Madhusudhan-et-al-2012-a}, but seem rare; main sequence dwarf carbon stars seem to be due to binary mass transfers and are unlikely sources for C-rich planets or ISOs \citep{Whitehouse-et-al-2018-a}.
If C-rich ISOs are discovered, they will very likely  be PMSOs, samples of the end of stellar evolution.

Over 0.2 M$_{\bigodot}$ of cold dust has been found in the ejecta of SN 1987A \citep{Indebetouw-et-al-2014-a}, 
and it seems possible that SN could create PMSOs in their remnants. However, 
the SN 1987A cold dust has an ejection velocity of at least 750 $\pm$ 250 km s$^{-1}$. If these velocities are typical of SN dust any SN PMSO would be either ejected from the galaxy or put into halo orbits, making a SN origin for 2I unlikely.
SN produced PMSO  would also be dominated by  
SN created elements, and so could be readily distinguished from s-process dominated AGB PMSO.

\section{2I/Borisov: An ISO from the Wolf 630 Dynamical Stream} 
\label{Sec:Wolf-630-Stellar-Stream}

Stellar perturbations make it hard to predict the detailed galactic trajectories of 
ISOs over intervals much longer than a few million years \citep{Zhang-2017-a}, 
$<$ 10$^{-3}$ of the history of the galaxy, and no firm association with a nearby star has been claimed for either 1I or 2I \citep{Bailer-Jones-et-al-2019-a,Hallatt-Wiegert-et-al-2019-a}.
It is thus unlikely that we will find the natal star system of any passing ISO; 
association with a stellar stream can serve as a proxy to provide at least some bounds on the age, metallicity, and chemical composition of the source stellar system. 

\begin{figure}[!ht]
\begin{center}
\includegraphics[scale=0.75]{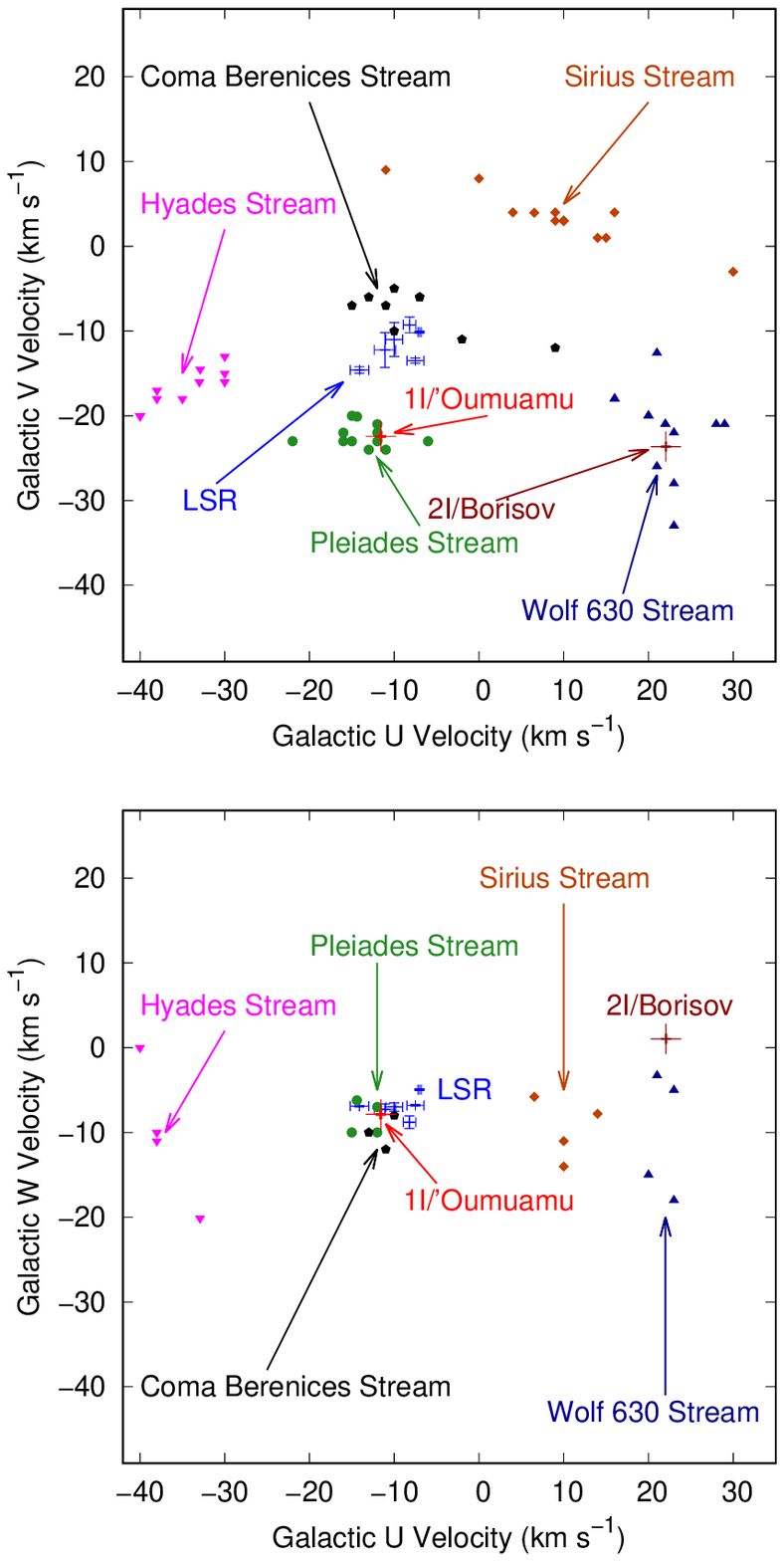}
\end{center}
\caption{
(Above) The galactocentric U and V components of velocity for 1I, 2I, the LSR and five local dynamical streams. The 1I and 2I incoming velocities are near the centroids of the determinations of the velocities of the Pleiades and Wolf 630 streams, respectively. The stream velocity estimates are from  \citep{Kushniruk-et-al-2017-a}, supplemented by \citep{Chereul-et-al-1998-b,Liang-et-al-2017-a,Gaia-et-al-2018-b}.
The 1I and LSR velocites are as shown in \citep{Eubanks-2019-b}, and the 2I velocity estimate is from the ephemeris of Bill Grey / Project Pluto.
(Below) The galactocentric V and W components of velocity; the data sources are as described above, with
fewer stream data points as not all surveys report the W velocity. The 2I incoming W velocity is near two of the determinations of Wolf 630 stream W velocity;  
the scatter between these velocity estimates may reflect substructure in the stream kinematics.
}
\label{fig:Velocity-Scatter}
\end{figure}

The Milky Way's velocity fields are highly non-uniform \citep{Ramos-et-al-2018-a},
with a substantial fraction of the stars in the solar neighborhood being concentrated 
in dynamical streams (also known as associations or moving groups) moving in the galaxy
(see, e.g., \citep{Kushniruk-et-al-2017-a,Gaia-et-al-2018-b}). 
Figure \ref{fig:Velocity-Scatter} shows  that the incoming 2I  velocity vector ``at infinity'' (\textbf{v}$_{\infty}$) 
is close to the motion of the Wolf 630 dynamical stream in all three velocity components, suggesting that 2I  was (before its interaction with the solar system) a member of that stream, just as 1I appears to have been a member of the Pleiades dynamical stream 
\citep{Feng-Jones-2018-a,Eubanks-2019-a,Eubanks-2019-b}. 

Connecting an ISO to a dynamical stream allows the stars
in the stream to be used as proxies for natal stellar system of the ISO.  
Data from a spectroscopic survey of stars with distances $<$ 250 pc
\citep{Quillen-et-al-2018-b} reveals that this stream 
possesses a slightly higher than solar metallicity ([Fe/H] $>$ 0.2 dex).
Antoja \textit{et al} 
\citep{Antoja-et-al-2008-a} determined the kinematics of 
over 16,000 nearby stars with known ages, the Wolf 630 
stream (their moving group 17) has in their data almost no stars younger than 
 0.5 Gyr, and are mostly 2 - 8 Gyr old. The lower bound of this range agrees well
 with an earlier estimate for the stellar age of the Wolf 630 stream of 2.7 $\pm$ 0.5 Gyr
 \citep{Bubar-King-2010-a}. C-rich AGB stars require roughly 0.5 - 2 Gyr to form; it seems likely that 
 Wolf 630 stars are old enough to form C-rich PMSO.

\section{Predictions}
\label{Sec:Predictions}

Spectroscopic observation of 2I  reveals  CN  (0-0)  gas  emission  \citep{Fitzsimmons-et-al-2019-a} and 
the [OI] oxygen line \citep{McKay-et-al-2019-a} but not water, OH or water ice \citep{Yang-et-al-2019-a}. 
Precovery observations (and cometary activity) have been found back to December 2018, when 2I was 7.8 AU from the Sun; this early activity is consistent with the presence of CO or CO$_{2}$ and in any case requires sublimation of substances more volatile than water ice  \citep{Ye-et-al-2019-c}. 
Although C$_{2}$ has been detected \citep{Lin-et-al-2019-a}, C$_{3}$ has not and 2I is regarded as a carbon chain depleted comet \citep{Opitom-et-al-2019-a,Kareta-et-al-2019-a}. All of these features are consistent with 2I being an PMSO object derived from C-rich stellar ejecta subject to strong stellar UV emissions during its formation.

%% An example table using AASTeX's deluxetable. Note that since
%% only one figure OR one table is allowed this is commented out.
\begin{deluxetable*}{lccc}
\tablecaption{Predictions for AGB/PNe PMSO\label{table:Predictions}}
\tablehead{
\colhead{Feature} & \colhead{O-rich} & \colhead{C-rich} & \colhead{N-rich} 
}
\startdata
Mass Ranges & M $\lesssim$  1.5 M$_{\bigodot}$ & 
1.5 M$_{\bigodot}$ $\lesssim$ M $\lesssim$ 3 M$_{\bigodot}$ &
M $\gtrsim$ 3.5 M$_{\bigodot}$ \\
Proportion & $\sim\frac{1}{4}$ & $\sim\frac{1}{2}$ &  $\sim\frac{1}{4}$ \\
C/O       &  $<$ 1  & $>$ 1, up to 3 & $\sim$0.05? \\
N/O       &  $\sim$0.3      & $\sim$0.3   & $\sim$1.3 \\
Water & Water Rich & Water Poor & Water Rich? \\
Dust & Al$_{2}$O$_{3}$ \& silicates & SiC \& Graphite. Si$_{3}$N$_{4}$? & Al$_{2}$O$_{3}$ \& silicates\\
Other &     & C aromatic bonds (PAH) & Greatly enhanced $^{7}$Li\\
     & & C aliphatic bonds      & Enhanced Na \\
$^{12}$C/$^{13}$C & $<$ 89 ? & $\sim$ 30 - 100  & $\sim$ 3.5 \\
$^{14}$N/$^{15}$N & $\sim$ 400 & $\sim$ 10$^{2}$ to 10$^{4}$ & $\sim$ 10$^{4}$ ? \\
$^{18}$O/$^{17}$O & $\sim$ 0.1 - 5  & ? &  $\sim$ 10$^{-4}$ \\
\enddata
\tablecomments{Predictions are derived by the author from results in \citep{Abia-et-al-2017-a,Busso-et-al-2014-a,Clement-et-al-2005-a,Davis-2011-a,Di_Criscienzo-et-al-2016-a,Kuchnet-Seager-2005-a,Ventura-et-al-2018-a,Ventura-et-al-2017-a,Wooden-et-al-2017-a}, and should be regarded as approximate.} 
\end{deluxetable*}  

A stardust comet with large amounts of volatiles
presumably resulted from molecular condensation, possibly from a cometary knot, and  
should be dominated by s-process elements, possibly dehydrogenated by
the intense radiation from PN stars.
Table \ref{table:Predictions} provides indications of the observational differences between the three classes of stardust chemistry,
based on observations of AGB and PNe stars, presolar grains and cometary dust.
CO is a very stable molecule, absorbing the C and O in the stellar envelope;
once the C/O ratio becomes $>$ 1, oxygen will be sequestered in CO and the ejected material will be largely free of oxides, including water. 
The failure to detect water suggests that 2I may derive from a C-rich stellar environment.

The estimated proportions in the second line in Table \ref{table:Predictions}, with the C-rich class being the most common, are from PNe data in 
\citep{Ventura-et-al-2017-a}.
C-rich comet dusts should be oxide poor, water poor, and rich in SiC, TiC and other carbides \citep{Kuchnet-Seager-2005-a}. Si$_{3}$N$_{4}$ may also be present \citep{Clement-et-al-2005-a}. AGB carbon dust has aromatic and aliphatic IR emissions at 3.28 and 3.4 $\mu$m, respectively \citep{Wooden-et-al-2017-a}, that may be present in a C-rich comet coma.
Finally, C-rich nebula have a number of unexplained spectral features that could be present in a PMSO spectrum, including potentially strong 21 $\mu$m and 30 $\mu$m emission lines \citep{Mishra-et-al-2016-a,Messenger-et-al-2013-a}.
N-rich PMSO  will have an excess of N, Na and possibly Li, and low amounts of $^{12}$C and $^{18}$O \citep{Di_Criscienzo-et-al-2016-a}.
Although O-rich stardust comets might appear to be compositionally similar to solar system comets, CNO
isotope ratios should vary, maybe by orders of magnitude from solar abundances. for any AGB PMSO 
\citep{Abia-et-al-2017-a,Davis-2011-a}, and should make it possible, if detected, to 
confirm the identity of PMSO of any chemistry. 

\section{Discussion and Conclusions} 
\label{Sec:Conclusions}

Roughly the same mass of metals is produced by post-main-sequence stars as is consumed by star formation, but only a fraction of the protostellar material is ejected into the InterStellar
Medium (ISM), while most or all of the AGB and SN ejecta is,
implying that even if PMSO production is substantially less efficient than proto-planetesimal production, there may be a greater flux of these objects in the ISM. 

Given the failure to detect water or any oxides at present, 2I seems be a possible C or N-rich 
stardust comet, probably dehydrogenated by PN UV radiation. 
I predict that 2I will continue to appear water poor; detection of 
the SiC emission line at $\sim$11.3 $\mu$m, or the IR Carbon lines, would provide a clear signature of a C-rich origin for 2I, which could be confirmed if the $^{12}$C/$^{13}$C ratio could be determined. 
The nature of 2I must be regarded at present as uncertain,
but it should be possible to determine definitely whether or not it is a PMSO with observations collected over the next few months.

The transit of ISOs through the solar system offers the opportunity to discover and study an entirely
new form of celestial body. Discovery and spectroscopic observation of PMSO would almost certainly lead to a much better understanding of post-Main Sequence stellar evolution. 
A PMSO comet could be observed at close range and sampled directly by \textit{in situ} spacecraft exploration \citep{Hein-et-al-2017-a,Hibberd-et-al-2019-b}, providing new insight into the formation of elements
essential for terrestrial biologies.

\acknowledgments
 The author was supported by Space Initiatives Inc and the Institute for Interstellar Studies-US (I4IS-US).
Thanks to Bill Gray of Project Pluto for sharing his 2I ephemerides. 

%%\appendix

%%\section*{References}

%%\begin{thebibliography}{}
%%end{thebibliography}


\bibliographystyle{aasjournal}
%%\bibliography{../eubanks}

\end{document}